\pdfoutput=1
\RequirePackage{ifpdf}
\ifpdf 
\documentclass[pdftex]{sigma}
\else
\documentclass{sigma}
\fi

\usepackage{cite}

\newcommand{\lalg}[1]{\mathfrak{#1}}  
\newcommand{\SU}{\mathrm{SU}}
\newcommand{\SO}{\mathrm{SO}}

\newcommand{\SL}{\mathrm{SL}}
\newcommand{\U}{\mathrm{U}}
\newcommand{\Spin}{\mathrm{Spin}}

\newcommand{\su}{\lalg{su}}

\newcommand{\so}{\lalg{so}}

\newcommand{\nb}{{\bf n}}
\newcommand{\mb}{{\bf m}}
\newcommand{\bb}{{\bf b}}

\newcommand{\tensor}{\otimes}
\newcommand{\hodge}{\ast}

\newcommand{\tens}{\otimes}

\newcommand{\ZZ}{\mathcal{Z}}

\newcommand{\dd}{\mathrm{d}}
\newcommand{\R}{\mathbb{R}}
\newcommand{\C}{\mathbb{C}}

\def\la{\langle}
\def\ra{\rangle}

\theoremstyle{definition}
\newtheorem{defi}{Definition}[section]
\theoremstyle{plain}
\newtheorem{lem}[defi]{Lemma}

\def\bra#1{\mathinner{\langle{#1}|}}
\def\ket#1{\mathinner{|{#1}\rangle}}
\def\braket#1#2{\mathinner{\langle{#1}|{#2}\rangle}}

\begin{document}

\allowdisplaybreaks

\renewcommand{\thefootnote}{$\star$}

\renewcommand{\PaperNumber}{008}

\FirstPageHeading

\ShortArticleName{The Construction of Spin Foam Vertex Amplitudes}

\ArticleName{The Construction of Spin Foam Vertex Amplitudes\footnote{This
paper is a~contribution to the Special Issue ``Loop Quantum Gravity and Cosmology''.
The full collection is available at \href{http://www.emis.de/journals/SIGMA/LQGC.html}{http://www.emis.de/journals/SIGMA/LQGC.html}}}

\Author{Eugenio BIANCHI~$^\dag$ and Frank HELLMANN~$^\ddag$}

\AuthorNameForHeading{E.~Bianchi and F.~Hellmann}

\Address{$^\dag$~Perimeter Institute for Theoretical Physics, Canada}
\EmailD{\href{mailto:ebianchi@perimeterinstitute.ca}{ebianchi@perimeterinstitute.ca}}

\Address{$^\ddag$~Max Planck Institute for Gravitational Physics (AEI), Germany}
\EmailD{\href{mailto:Frank.Hellmann@aei.mpg.de}{Frank.Hellmann@aei.mpg.de}}

\ArticleDates{Received July 20, 2012, in f\/inal form January 27, 2013; Published online January 31, 2013}

\Abstract{Spin foam vertex amplitudes are the key ingredient of spin foam models for quantum gravity. These fall into the realm of discretized path integral, and can be seen as generalized lattice gauge theories. They can be seen as an attempt at a 4-dimensional generalization of the Ponzano--Regge model for 3d quantum gravity. We motivate and review the construction of the vertex amplitudes of recent spin foam models, giving two dif\/ferent and complementary perspectives of this construction. The f\/irst proceeds by extracting geometric conf\/igurations from a topological theory of the BF type, and can be seen to be in the tradition of the work of Barrett, Crane, Freidel and Krasnov. The second keeps closer contact to the structure of Loop Quantum Gravity and tries to identify an appropriate set of constraints to def\/ine a Lorentz-invariant interaction of its quanta of space. This approach is in the tradition of the work of Smolin, Markopoulous, Engle, Pereira, Rovelli and Livine.}


\Classification{81T25; 81T45}


 \renewcommand{\thefootnote}{\arabic{footnote}}
\setcounter{footnote}{0}

\section{Introduction}
A quantum theory of gravity is expected to provide a def\/inition of the Feynman integral over gravitational degrees of freedom (d.o.f.), weighted with the exponential of the gravity action,
\begin{gather}\label{eq-GravityPathIntegral}
Z=\int_{\text{Gravitational d.o.f.}} \hspace{-4em}e^{\frac{i}{\hbar} \displaystyle S_{\text{Gravity}}} .
\end{gather}

The spin foam approach \cite{Reisenberger1997a,Reisenberger1997b,Baez1998b,Freidel1999a} is among those approaches that attempt to do so by f\/irst substituting the formal expression above by a def\/inite integral over a f\/inite number of degrees of freedom. As opposed to approaches like (causal) dynamical triangulations \cite{Ambjorn1992,Ambjorn2010}, or quantum Regge calculus \cite{Rocek1981,Williamsd}, the discretized models are not arrived at by a direct discretisation of second order gravity\footnote{Though a derivation along these lines might be possible, if in a somewhat ad hoc way, see \cite{Baratin:2008du}.}. Instead f\/irst order formulations of gravity more closely related to gauge theory, in particular those related closely to classical BF theory, are used. It can be seen as an attempt at lattice gauge gravity. We will brief\/ly review these constructions in Section~\ref{sec-ClassicalPreliminaries}.

This has various advantages. For one, it is possible to make contact with the Hilbert space of Loop Quantum Gravity \cite{Rovelli:2004tv,Ashtekar:2004eh,Thiemann:book2007}, which is also based on gauge theoretic formulations of general relativity. This means that the geometric degrees of freedom are genuinely quantum mechanical objects here. Furthermore one can exploit gauge theoretic topological quantum f\/ield theories (TQFTs) in the construction of the discretised theory. The most pertinent is quantum $\SU(2)$ BF theory which is reviewed in Section~\ref{sec-quantumBF}. There we also recall how the geometric degrees of freedom are contained in the BF degrees of freedom. Quantum deformed TQFTs can also serve as a starting point~\cite{Fairbairn2010,Han:2010pz}.

These TQFTs are def\/ined on a discrete structure, namely a 2-complex (or a branched surface, made of faces, edges and vertices) embedded in the manifold. This mirrors the situation in lattice gauge theory, where a regular lattice with 2-dimensional plaquettes f\/ills spacetime. The topological nature of the TQFTs can be seen in the fact that they are independent of this discrete structure. While these are not well def\/ined at the level of the partition function, and can possibly not be regularized just on a 2-complex~\cite{Bonzom:2012mb,Bonzom:2010zh,Bonzom:2011br,Bonzom:2012mb,Bonzom:2010ar,Baratin:2011tg}, see~\cite{Bahr-toAppear} though, this has no bearing on the construction of the vertex amplitude itself.

The geometric interpretation of BF theory is most apparent in a dual picture. There the action is not primarily discretized on the plaquettes of the discrete structure, as in ordinary lattice gauge theory, but is instead localized to the vertices\footnote{For other ways to formulate these models, we refer the reader to~\cite{DCT,Bahr2011,Perez2011} for other examples.}.

This is the origin of the term vertex amplitude in the title of this paper. These vertex amplitudes depend on group representation theoretic data labeling the edges and faces neighbouring the vertex. These have a clear geometric meaning if the 2-complex is dual to a triangulation $\Delta$ in a sense explained in Section~\ref{sec-quantumBF}. In particular we have representations~$j_f$ on the faces of the 2-complex and a basis of intertwiners on the edges. They are denoted $\iota_e \in \operatorname{Inv}(j_f \tens j_{f'} \cdots)$ and live in the invariant subspace of the tensor product of the representations $j_f$ labeling the faces that have $e$ in its boundary. The vertex amplitude is then given by
\begin{gather*}
  A_v=\big\{\!\mathop{\otimes}\iota_e\big\},
\end{gather*}

where $\{\}$ denotes the natural contraction of all intertwiners $\iota_e$ sharing a face, using invariant inner products on the representation spaces $j_f$. Details of this construction are given in Section~\ref{sec-quantumBF}.

A discretisation of~\eqref{eq-GravityPathIntegral} is obtained by restricting the variables of the TQFT on the discrete structure to the geometric ones contained in the theory already. This breaks independence from the discrete structure. Heuristically we want to think of this discrete theory as a path integral where only a discrete subset of the gravitation degrees of freedom are switched on. We will show in Section~\ref{sec-GeometricConstruction} how this can be achieved. Remarkably, this can be done while also keeping the relationship with the Loop Quantum Gravity state space by respecting the group representation theoretic nature of the quantum BF degrees of freedom. The restriction of degrees of freedom for that case can be formulated by writing
\begin{gather*}
 A_v=\big\{\!\mathop{\otimes}\mathcal{I}_\gamma(i_e)\big\},
 \end{gather*}
where $\mathcal{I}_\gamma(i_e)$ now is a map that parametrises a subspace of the $\Spin(4)$ or $\SL(2,\C)$ BF degrees of freedom, that is, the intertwiner spaces, in terms of $\SU(2)$ intertwiners. This subspace is then intended to be the subspace of geometric degrees of freedom in the BF theory.

These restrictions to geometric degrees of freedom can be done by identifying the geometric content of the intertwiner spaces directly \cite{Reisenberger1997b,Barrett:1997gw,Barrett2000,Freidel:2007py} or by attempting to discretize and quantize a set of so called simplicity constraints $C(B)=0$ on the $B$-f\/ield that reduce classical BF theory to a theory containing General Relativity~(GR). Several dif\/ferent sets of constraints along these lines are known, and these lead to the slightly dif\/ferent models of Freidel and Krasnov~\cite{Freidel:2007py, Livine:2007ya}, Engle, Pereira, Livine and Rovelli \cite{Engle:2007qf,Engle:2007uq,Engle:2008ev,Engle:2007wy,Engle2011e}, Dupuis and Livine~\cite{Dupuis:2010iq,Dupuis:2011fz,Dupuis:2011dh,Dupuis:2011wy} and Baratin and Oriti~\cite{Baratin:2010wi,Baratin:2011tx,Baratin:2011hp}. This leads to the interpretation of the vertex amplitude as a functional integral for a BF theory, with constrained $B$-f\/ields on the $2$-skeleton $\Delta_2$ of the triangulation $\Delta$,
\begin{gather*}
A_v=\int_{C(B)=0  \  \text{on}\   \Delta_2} \mathcal{D}B \mathcal{D}\omega  e^{i S_{\rm BF} [B,\omega]}.
\end{gather*}

The construction outlined in Sections \ref{sec-quantumBF} and~\ref{sec-GeometricConstruction} follows the f\/irst option, and directly identif\/ies the geometric content of the intertwiner spaces. It relies on the specif\/ic form of the 2-complex, as various geometricity results fail for arbitrary complexes. In order to have a complete matching with the Loop Quantum Gravity Hilbert space such a restriction on the combinatorial data needs to be lifted.

This can be done, taking as inputs the constraint formulation of geometricity, and is demonstrated in Section~\ref{sec-Principles}. There a set of principles are given in which one can understand the spin foam amplitude in a more general setting than the simplicial one. However, in this generalisation the geometricity results underlying the preceding constructions are lost.

We conclude by discussing the trade of\/fs between various constructions and several open issues in Section~\ref{sec-OpenIssues}.

Note that we do not attempt a historic overview of the developments, but rather try to present somewhat novel perspectives on the construction and motivation of the currently most analyzed models. In particular we will omit the Barrett--Crane model, as we merely aim to ref\/lect the activities and results of the last roughly f\/ive years. For a very complete overview of the entire f\/ield we refer the interested reader to the excellent recent review~\cite{Perez2011}.

\section{Classical preliminaries}\label{sec-ClassicalPreliminaries}

Spin foam models are motivated from the classical form of Euclidean and Lorentzian general relativity written in terms of gauge theoretic variables. We will very brief\/ly review the relevant theories here. The fundamental f\/ields will be an $\R^4$ valued $1$-form $e^I$, a skew-symmetric $\R^4 \times \R^4$ valued $2$-form $B^{IJ}$ and an $\SO(4)$ or $\SO^{\uparrow}(3,1)$ connection $\omega^{IJ}$, where $I, J = 0, \dots, 3$. Indices are raised and lowered using the Euclidean or Minkowskian metric on $\R^4$, thus, if it is non-degenerate, $e^I\otimes e_I$ is a Euclidean or Lorentzian metric on our manifold. The action of gravity can be written in terms of the f\/ields $e$ and $\omega$ as\footnote{In the following, we work in units $16\pi G=1$, $c=1$, $\hbar=1$.}
\begin{gather} \label{eq-HolAction}
S_{\rm GR}[e,\omega] =  \frac{1}{2}\int \epsilon_{IJKL} e^I \wedge e^J \wedge F^{KL}(\omega)  +  \frac{1}{\gamma} \int e_I \wedge e_J \wedge F^{IJ}(\omega).
\end{gather}
This is the Einstein--Cartan action for general relativity,
supplemented with the Holst term~\cite{Holst1996a}. This additional term $\frac{1}{\gamma} \int e_I \wedge e_J \wedge F^{IJ}(\omega)$, does not contribute to the classical equations of motion but changes the quantum theory. The coupling parameter $\gamma$, called the Immirzi parameter, thus gives a one parameter family of quantization ambiguities.

Using $B^{IJ}$ and $\omega^{IJ}$ we can write an action for a theory aptly named BF theory \cite{Horowitz:1989ng}
\begin{gather} \label{eq-BFAction}
S_{\rm BF}[B,\omega] =  \int_M B_{IJ} \wedge F(\omega)^{IJ}.
\end{gather}
The equations of motion for BF theory are much stronger than those of GR. Stationarity of the action with respect to the  $B$-f\/ield, imposes that the classical solution have connection $\omega$ that is locally f\/lat
\[
F(\omega)=0 .
\]
As a result, the theory has no local degrees of freedom. On the other hand, if the $B$-f\/ield is constrained to be of the form
\begin{gather}
B^{IJ}=\frac{1}{2}{\epsilon^{IJ}}_{KL}  e^K \wedge e^L + \frac{1}{\gamma} e^I \wedge e^J ,
\label{eq:simpleB}
\end{gather}
local degrees of freedom are allowed, and the dynamics reduces to the one of general relativity.

This observation gives rise to the so-called Pleba\'{n}ski formulation of general relativity
\begin{gather*}  
S_{\rm CJDM} = \int_M B_{IJ} \wedge F(\omega)^{IJ} + \lambda C(B),
\end{gather*}
where $\lambda$ is a Lagrange multiplier and $C(B)$ refers to a set of constraints that forces~$B$ to be of the form (\ref{eq:simpleB}). The constraints $C(B)$ are usually called the simplicity constraints in the context of spin foams.

This action was f\/irst considered by Capovilla, Jacobson, Dell and Mason in~\cite{Capovilla:1989ac,Capovilla:1991qb}, and is based on the self-dual decomposition of the $B$-f\/ield going back to Pleba\'{n}ski's work~\cite{Plebanski:1977zz}, see~\cite{Krasnov:2009pu} for a modern introduction. Another method of constraining the theory appropriately is given in~\cite{Reisenberger1998}. It should be noted that both~\cite{Capovilla:1989ac} and~\cite{Reisenberger1998} point out that the chosen constraints do not completely reduce the theory to gravity, but that a number of pathological solutions remain. For further details on the relationship between BF theory and gravity see also the review~\cite{Freidel:2012np} in the same special issue.

\section{Quantum BF theory}\label{sec-quantumBF}

The underlying structure, and prototypical example of all new vertex constructions \cite{Freidel:2007py,Engle:2007qf,Engle:2007uq,Engle:2007wy,Engle:2008ev,Fairbairn2010,Dupuis:2010iq,Kaminski2010a,Engle2011e,Dupuis:2011fz,Dupuis:2011dh,Dupuis:2011wy} is the vertex amplitude of a lattice quantization of BF theory. This lattice quantization was f\/irst described in arbitrary dimensions by Ooguri in \cite{Ooguri1992c}. In three dimensions it is the well known Ponzano--Regge state sum model~\cite{Barrett:2008wh,ponzanoregge}. Heuristically the idea is to integrate out the $B$-f\/ield. It then acts as a Lagrange multiplier enforcing the f\/latness of the connection. The path integral then reduces to an integral over f\/lat connections.

This integral is implemented in a discretized fashion by integrating connections along the edges of the dual of a triangulation of the manifold, that is, we associate a group element $g_e$ to every such edge. Flatness is then implemented by enforcing the oriented product of group elements around each face, $H_f(\{g_e\}_f) = g_e^{\pm1} \cdots g_{e'}^{\pm1}$, of the dual triangulation to be the identity. Implementing this condition with delta function we obtain an integral over the moduli space of f\/lat connections. Up to problems of regularisation\footnote{Note that these are not well understood in general, and it is actually challenging to make this formulation precise. See~\cite{Bonzom:2010ar,Bonzom:2010zh,Bonzom:2011br,Bonzom:2012mb}.} this is independent of the triangulation chosen, and thus captures the continuum path integral completely:
\[
\ZZ_{\sigma} =\int \left( \prod_e \dd g_{e} \right) \left( \prod_f H_f(\{g_e\}_f) \right).
\]

The delta functions implementing the f\/latness can be developed into characters, and the group integrations performed exactly. The result is a labeling of the faces of the dual triangulation by irreducible representations, and of the dual edges by intertwiners between these representations. The intertwiners are then contracted at the vertices.

More precisely we can consider combinatorial $2$-complexes,  $\sigma=\{f,e,v\}$ consisting of a~collection of elements called faces $f$, edges $e$, and vertices $v$, together with boundary relations that associate an ordered set of dual edges to the boundary of a dual face, $\partial f= \{e_1,e_2, e_3,\ldots\}$, and an ordered couple of vertices to the boundary of an edge, $\partial e= \{v_1,v_2\}$. Moreover, we say that an edge belongs to a vertex, $e\in v$, if $v\subset \partial e$. A spin foam conf\/iguration consists in an assignment of a spin $j_f$ to each face of the $2$-complex, and an intertwiner~$\iota_e$ between the representations $j_f$ with $e \in f$ to each edge. The full state sum is then given by
\begin{gather*}
\ZZ_\sigma =\sum_{j_f,\iota_e} \prod_{f\subset \sigma}(2j_f+1) \prod_{v\subset \sigma}\big\{\!\mathop{\otimes}_{e\in v}  \iota_e\big\} ,
\end{gather*}
where, as mentioned in the introduction, $\{~\}$ denotes the contraction of indices belonging to the same face. In order to def\/ine the contraction using hermitian inner products an ordering is needed, this can be taken from the orientation of the face. Alternatively a graded antisymmetric bilinear inner product can be used for which the ordering merely contributes an overall sign factor.

\begin{figure}[t]
\centering
\includegraphics[scale=0.9]{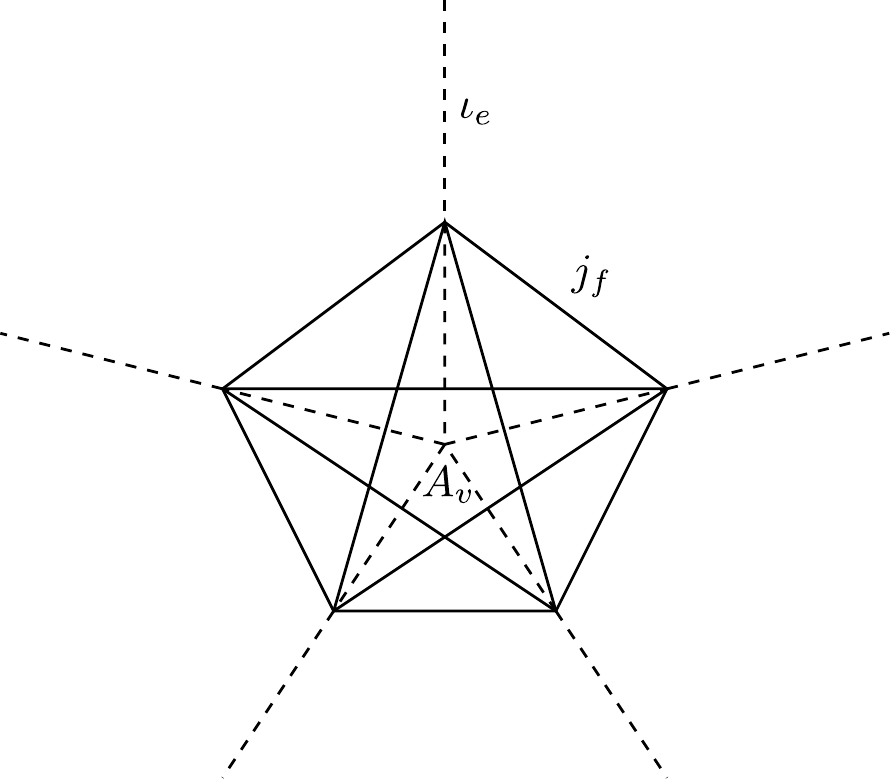}
\caption{Vertex amplitude combinatorics. The dashed lines indicate the edges of the 2-complex, meeting at a vertex in the center, and carrying an intertwiner~$\iota_e$. Every triangle formed from a solid line and two dashed line lies in a face of the 2-complex and caries a representation~$j_f$. The solid lines on their own indicate the pattern of contraction among the intertwiners as given in equation~\eqref{eq-ContractionPattern}, def\/ining the vertex amplitude~$A_v$.}
\label{fig-15j}
\end{figure}

Consider now again the case of a 2-complex dual to a triangulation. As the intertwiners live on dual edges, they are associated to $(n-1)$-simplices in the triangulation. Contractions happen along the faces of the dual triangulation and thus correspond to an $(n-2)$-simplex. The pattern of contraction is the dual of the surface of an $n$-simplex, which is again an $n$-simplex. As mentioned in the introduction, the term vertex in vertex amplitude, refers to the dual to the triangulation, at the level of the triangulation it is the amplitude associated to the $n$-simplices. In four dimensions we have explicitly
\begin{gather}\label{eq-ContractionPattern}
 A_v(\iota) = \big\{\!\mathop{\otimes}_{e\in v}\!\iota_e\big\} = \iota_1^{(12)(13)(14)(15)}\! \iota_2^{(12)(23)(24)(25)}\! \iota_3^{(13)(23)(34)(35)} \! \iota_4^{(14)(24)(34)(45)}\! \iota_5^{(15)(25)(35)(45)}.\!\!
\end{gather}
Here $(12)$ indicates an index associated to the face between the edges $1$ and $2$. The pattern of contractions being the same as in Fig.~\ref{fig-15j}.

\section{Spin foams from BF theory with geometricity conditions}\label{sec-GeometricConstruction}

Let us now specify to a $4$-dimensional $\SU(2)$ BF theory. Our strategy will be to identify the geometric asymptotics of this amplitude in order to understand how to identify the geometric subsector of the theory. This will of course only give a semiclassical criterium for the reduction of the BF amplitudes to geometric ones, and we will indeed see dif\/ferent quantum mechanical constructions satisfying it, including the representation theoretic one of Engle, Pereira, Rovelli and Livine, and the holomorphic one of Dupuis and Livine, as well as the one based on quantisation of the $B$-f\/ield by Baratin and Oriti. We will give the resulting amplitude in Section~\ref{sec:Furthers}.

Here we will take the vertex amplitude~\eqref{eq-ContractionPattern} as our starting point. The intertwiners contracted there are dual to tetrahedra. To bring out the geometry in the construction it will be convenient to write the intertwiners in a coherent state basis, that is, in terms of states that are the eigenvectors of hermitian $\su(2)$ generators (see Appendix~\ref{app:reps}). Let $\nb \cdot L$ be the generator of rotations around the axis~$\nb$, then we call $|\nb\ra_j$ its heighest weight eigenvectors in the~$j$ representation
\begin{gather*}
\nb \cdot L |\nb\ra_j = j |\nb\ra_j.
\end{gather*}

We then obtain the coherent intertwiners of \cite{Livine:2007vk}
\begin{gather}\label{eq-coherentInts}
\iota_1 (\nb) = \int_{\SU(2)} \dd g\; g|\nb_{12}\ra_{j_{12}} \tens g|\nb_{13}\ra_{j_{13}} \tens g|\nb_{14}\ra_{j_{14}} \tens g|\nb_{15}\ra_{j_{15}}.
\end{gather}

These are specif\/ied by the vectors $\nb$ up to an overall phase only. This, however, cancels in the complete state sum and will not play a role in what is to follow.

Similarly for the other intertwiners, where now, taking $a,b \in 1, \dots, 5$, we have $\nb_{ab} \neq \nb_{ab}$, but $j_{ab} = j_{ba}$. The vertex amplitude then takes the form
\begin{gather*}
 A(j,\nb) = \pm \prod_{c} \int_{\SU(2)} \dd g_c \prod_{a < b} \la - \nb_{ab}| g_a^{-1} g_b|\nb_{ba}\ra_{j_{ab}}.
\end{gather*}

Here the minus sign in front of $\nb_{ab}$ occurs because we use the epsilon inner product for contraction. This is bilinear but graded anti-symmetric. Thus the orientation of edges inf\/luences the amplitude only through a sign, which we will disregard here.

Each vector and spin label is associated to a $(n-2)$-simplex, that is, a triangle $\Delta_{ab}$. It is natural to associate the $j \nb$ with the bivector f\/ield $B$ of the continuum action, which is always an $(n-2)$-form that can naturally be integrated against the (oriented) $(n-2)$-simplices. This is borne out by the asymptotic geometry of the amplitude. Scaling the spin labels to $\lambda j$ the leading order behaviour of the amplitude in inverse powers of $\lambda$ is polynomial rather than exponentially suppressed if there exist rotations $g_a$ such that
\begin{gather*}
 \bb_{ab} = j_{ab}   g_{a}   \nb_{ab}
\end{gather*}
satisfy
 \begin{gather}\label{eq-ori}
  \bb_{ab} = -\bb_{ba}
\end{gather}
 and
 \begin{gather} \label{eq-clo}
 \sum_{b :\; b\neq a} \bb_{ab}= 0.
 \end{gather}
 Now it is easy to see (Section~2.2.1 in \cite{Barrett:2009as}, or Lemma~4.2.1 in \cite{Hellmann:2010nf}) that these conf\/igurations correspond to simplexwise constant $\su(2)$-valued bivector f\/ields. Schematically the correspondence is given by setting
\begin{gather*}
\bb_{ab} = \int_{\Delta_{ab}} B.
\end{gather*}

Thus we see that, doubling the group to $\SU(2) \times \SU(2) \ni \SO(4)$, the intertwiners are the natural place to attempt to introduce the restriction to a geometric sector in the theory. Note however that whereas the BF theory captures the whole of the dynamics of the continuum theory on the lattice, this will no longer be the case here, as the restrictions are only implemented on the 3-dimensional hypersurfaces of the lattice, which is thus no longer f\/iducial. The interior of the 4-simplices continues not be described by BF. Thus the 3-dimensional hypersurfaces are now singled out and become visible, the independence from the lattice is broken.

\subsection{Euclidean theory}

Among the simplexwise constant $\su(2)$ valued $B$-f\/ields the geometric ones are distinguished by the following result (e.g.\ from Theorem~4.2.5 in\cite{Hellmann:2010nf}):

\begin{lem}[geometric $B$-f\/ields]
If and only if $\tilde{\nb}_{ab} = j_{ab} \nb_{ab}$ are the area normals of the tetrahedra in the boundary of a non-degenerate, geometric Euclidean $4$-simplex, are there two inequivalent solutions $\bb^+$ and $\bb^-$ to the critical point equations~\eqref{eq-ori} and~\eqref{eq-clo}.
\end{lem}

The appearance of a second solution can be understood from Hodge duality. The space of bivectors in $\R^4$ is six-dimensional and can be identif\/ied with the generators of four-dimensional rotations, that is, with $\so(4)$. As $\so(4)$ is $\su(2) \times \su(2)$ the space of bivectors decomposes into a left and right part. That is, we can choose generators $L^\pm$ that satisfy
\begin{gather*}
  \frac12 (1 \pm \hodge) L^\pm = L^\pm,
\end{gather*}
and
\begin{gather*}
  \frac12 (1 \mp \hodge) L^\pm = 0.
\end{gather*}
Together the left and the right part then form the bivectors of the geometric 4-simplex in question, that is
\begin{gather} \label{eq-geomBiv}
\bb^+ \cdot L^+ + \bb^- \cdot L^- = \hodge (e \wedge e)_{\Delta}.
\end{gather}

The subscript $\Delta$ here means that the constant two form $e \wedge e$ is integrated over the triangle to give a Lie algebra element, that is, with appropriate orientations and~$e_1$ and~$e_2$ being two edge vectors of the triangle $\Delta$ we have
\begin{gather*}
(e \wedge e)^{IK}_{\Delta} = e^{[I}_1 e^{K]}_2.
\end{gather*}

The left and right part individually code the 3-dimensional boundary geometry. To see this note that
for a tetrahedron in the 4-simplex that is orthogonal to the north pole we can def\/ine
\begin{gather*}
\pm(0, \tilde{\nb})^J = \frac12 (1,0,0,0)^I [(1 \pm \hodge) (e \wedge e)_{\Delta}]^{IJ} = \pm \frac12 \epsilon^{0JKL} e_1^K e_2^L.
\end{gather*}

Geometrically $\tilde{\nb}$ is the area normal to the tetrahedron in $\R^3$. As these code the tetrahedral geometry, the full boundary geometry is contained in each sector of the theory. As the Hodge operator is invariant under rotations, so is the split into left and right sector. Thus if the face is not orthogonal to the north pole we can rotate it there and read of the three-dimensional data in the same way. As $\SO(4)$ acts on the left and right separately, it follows that the geometry def\/ined by these outward area normals is indeed the geometry of the surface of the 4-simplex.

On the other hand the bivectors satisfy closure~\eqref{eq-clo} and orientation~\eqref{eq-ori}, and as these are linear equations, they do so in the left and right sector independently. Thus we see that if we have data that corresponds to a geometric boundary, we immediately obtain two solutions from the left and right sector of the bivectors of the geometric 4-simplex. It can also be shown by considering the area normals, that these need to be two genuinely independent solutions.
This demonstrates that in the geometric sector we have indeed two solutions. The reverse implication is somewhat harder to demonstrate, and we refer the reader to the literature~\cite{Barrett2009,Barrett:2009as,Hellmann:2010nf}.

In order to add a geometric sector above the $B$-f\/ield conf\/igurations we can use a simple trick. Doubling the group to $\Spin(4) = \SU(2) \times \SU(2)$ we have two sets of data~$j^\pm$ and~$\nb^\pm$. Then picking $j^+ = j^-$ and $\nb^+ = \nb^-$ simply means for generic conf\/igurations we obtain one solution to the critical point equations
\begin{gather*}
 b_{ab}^{\pm} = j_{ab} g_{a} \nb_{ab}.
\end{gather*}

However, in the geometric sector we have two solutions of the underlying theory which we denote $b^{1/2}$. That means there are four solutions to the critical point equations
\begin{gather*}
 (b^+, b^-) \in \big\{\big(b^1, b^1\big),\big(b^1, b^2\big),\big(b^2, b^1\big),\big(b^2, b^2\big)\big\}.
\end{gather*}

The diagonal terms mirror the $\SU(2)$ BF solutions, thus, after doubling and identifying we retain the full $\SU(2)$ BF theory. However, in addition we have gained two more asymptotic solutions corresponding to a full geometric bivector~\eqref{eq-geomBiv}. Thus we have added a geometric sector to the theory.

The construction by identifying the left and the right part of a $\Spin(4)$ BF theory was f\/irst suggested in the quantum context in~\cite{Livine:2007ya} and~\cite{Freidel:2007py}. The classical idea goes back to~\cite{Reisenberger1998}. This is, however, not the only way of implementing a geometric sector. For this it is suf\/f\/icient to map the $\su(2)$ data specifying the boundary geometry to $\su(2) \times \su(2)$ in such a way that the asymptotic geometry is preserved, that is, such that the left and right sector agree on the geometry induced on the boundary. Furthermore the underlying mechanism is not af\/fected by choosing to identify the left and the right sector through $j^+ = c j^-$ for some f\/ixed $c$ and/or $\nb^+ = - \nb^-$, which provides us with a large variety of models. These choices have a correspondence in the continuum theory in the Holst action where we have $B = (\hodge + \gamma)(e \wedge e),$ with $c = c_{\gamma} = \frac{1+\gamma}{|1-\gamma|}$.

One of the aims of the construction in \cite{Engle:2007uq,Engle:2007wy,Engle:2007qf,Pereira:2007nh} was to have a pure $\SU(2)$ boundary state space. That is, the doubling and the reduction of the theory should not change the boundary Hilbert space away from the $\SU(2)$ BF one, as the $\SU(2)$ BF Hilbert space coincides with the loop quantum gravity one. This can be achieved by choosing a more general map from $\su(2)$ to $\su(2) \times \su(2)$ that doesn't just map the coherent state data, but the actual representation spaces. The idea is to introduce a representation $k$ such that $j^\pm = \frac{|1 \pm \gamma|}{2} k$, and map the $k$ representation space to the lowest or highest weight subspace, which is just this $k$, in
\begin{gather*}
 V(j^+) \tens V(j^-) = \bigoplus_{\frac12 (|1 - \gamma| - |1 + \gamma|) k}^{\frac12 (|1 - \gamma| + |1 + \gamma|) k} V(k'),
\end{gather*}
by the natural inclusion map. For $\gamma < 1$ this leads to the same construction as the direct identif\/ication of coherent state data outlined in the preceding paragraph, whereas for $\gamma > 1$ we incur additional integrations
\begin{gather*}
I_{\rm EPRL} (|\nb_{1b}\ra_{k_{1b}})  =  \prod_{b\neq 1} \int_{S^2 \times S^2}\!\! \dd \mb_{1b}^+\, \dd \mb_{1b}^- \, \la \mb^+_{1b}|\nb_{1b}\ra_{k_{1b}} \la- \mb^+_{1b}| \mb^-_{1b}\ra_{j^-_{ab}}   \bigotimes_{b \neq 1} |\mb^+_{1b}\ra_{j^+_{ab}} \tens |\mb^-_{1b}\ra_{j^-_{ab}}.
\end{gather*}

These maps retain the asymptotic geometry required to give a geometric sector in the state sum, but now also have a straightforward def\/inition in terms of representation theory. In this way the map is uniquely specif\/ied upon the choice of $j^+$, $j^-$ and $k$. We can use these maps to def\/ine and parametrise what should represent the geometric subspace of the $\Spin(4)$ intertwiner space by constructing
\begin{gather*} 
\mathcal{I} = \int_{\Spin(4)} \dd g \;  g I_{\rm EPRL} \tens g I_{\rm EPRL} \tens g I_{\rm EPRL} \tens g I_{\rm EPRL}.
\end{gather*}

There are some ambiguities that arise in this parametrisation. As pointed out in \cite{Kaminski2010,Kaminski2010a,Kaminski2011}, the $\SU(2)$ parametrisation can fail to be injective, and the inner product induced by it dif\/fers from the natural $\SO(4)$ invariant one. These issues can be f\/ixed only at the price of giving up the matching to the Loop Quantum Gravity Hilbert space.

Another way to think of these injection maps is to consider the generators of $\Spin(4)$ and its $\SU(2)$ subgroup, for this it will be convenient to change the basis of the Lie algebra slightly. We have begun with $L^\pm$ satisfying
\begin{gather*}
[L^\pm_i, L^\pm_j]  = i \epsilon^{ijk} L^\pm_k,\qquad
[L^+_i, L^-_j]  = 0.
\end{gather*}

The doubling of the left and right data means simply that we choose simultaneous eigenstates of the two commuting operators $\nb \cdot L^\pm$ to construct the amplitude. That is, we used coherent simple bivectors.

Now to bring out the interpretation in terms of representation theory it is convenient to pick a dif\/ferent basis
\begin{gather*}
  L^\pm_{i}  =  \frac12(L_i \pm K_i),\qquad
L_{i}  =  L^+_i + L^-_i,\qquad
K_{i}  =  L^+_i - L^-_i,
\end{gather*}
in terms of which we have
\begin{gather*}
  [K_i,K_j]  =  i \epsilon^{ijk} L_k,\qquad
 [K_i, L_j]  =  i \epsilon^{ijk} K_k,\qquad
 [L_i, L_j]  =  i \epsilon^{ijk} L_k.
\end{gather*}

We see that the $L_i$ are the generators of the diagonal subgroup. The~$K$ do not generate a~subgroup but parametrize the coset space of the diagonal subgroup $\Spin(4)/\SU(2)_d = S^3$. The parameter $\gamma$ relates the highest weight eigenvalues of the generators in the same direction. To see this consider the action on a coherent bivector for $\gamma < 1$
\begin{gather*}
  \nb \cdot L |\nb\ra_{\frac12(1+\gamma)k} \tensor |\nb\ra_{\frac12(1-\gamma)k} = k |\nb\ra_{\frac12(1+\gamma)k} \tensor |\nb\ra_{\frac12(1-\gamma)k},
\end{gather*}
while
\begin{gather*}
  \nb \cdot K |\nb\ra_{\frac12(1+\gamma)k} \tensor |\nb\ra_{\frac12(1-\gamma)k} = \gamma k |\nb\ra_{\frac12(1+\gamma)k} \tensor |\nb\ra_{\frac12(1-\gamma)k} . \end{gather*}

That is, on coherent bivectors we have $\gamma \nb \cdot L \ket{\rm biv} = \nb \cdot K \ket{\rm biv}$. This condition can be seen as a discrete implementation of the constraints $C(B)$, and will be a key ingredient in the construction in Section~\ref{sec-Principles}.

\subsubsection{Summary}

Putting the above together we arrive at the following possibilities for the Euclidean vertex amplitudes.
The simple doubling and unbalancing with $\nb^+ = \nb^-$ and $j^+ = c_\gamma j^-$ leads to
\begin{gather*}
 A_v  = e^{i\phi} \int_{\SU(2)^{10}} \left(\prod_{c=1,\dots,5} \dd g_c \dd g'_c\right)
  \prod_{a<b}\bra{-\nb_{ab}} g_a^{-1} g_b \ket{\nb_{ba}}_{ j_{ab}} \bra{- \nb_{ab}} {g'}_a^{-1} g'_b \ket{\nb_{ba}}_{ c_\gamma j_{ab}}.
\end{gather*}
Here $\phi$ is the undetermined phase and sign factor. The doubling with sign and unbalancing with $- \nb^+ = \nb^-$ and $j^+ = c_\gamma j^-$ leads to
\begin{gather*}
 A_v  = e^{i\phi} \int_{\SU(2)^{10}} \left(\prod_{c=1,\dots,5} \dd g_c \dd g'_c\right)   \prod_{a<b} \bra{-\nb_{ab}} g_a^{-1} g_b \ket{\nb_{ba}}_{ j_{ab}} \bra{\nb_{ab}} {g'}_a^{-1} g'_b \ket{- \nb_{ba}}_{ c_\gamma j_{ab}}.
\end{gather*}
which using $\bra{-\nb} g \ket{-\nb'} = \overline{\bra{\nb} g \ket{\nb'}}$ takes the form of the model of Freidel and Krasnov \cite{Freidel:2007py}
\begin{gather*}
 A_v  = e^{i\phi} \int_{\SU(2)^{10}} \left(\prod_{c=1,\dots,5} \dd g_c \dd g'_c\right)   \prod_{a<b} \bra{-\nb_{ab}} g_a^{-1} g_b \ket{\nb_{ba}}_{j_{ab}} \overline{\bra{- \nb_{ab}} {g'}_a^{-1} g'_b \ket{\nb_{ba}}}_{ c_\gamma j_{ab}}.
\end{gather*}

Finally the representation theoretic model of Engle, Pereira, Rovelli and Livine \cite{Engle:2007uq,Engle:2007wy,Engle:2007qf,Pereira:2007nh}, def\/ined using $I_{\rm EPRL}$ takes the form
\begin{gather*}
 A_v = e^{i\phi} \int_{\SU(2)^{10}} \left(\prod_{c=1,\dots,5} dg_c dg'_c\right) \int_{{S^2}^{20}} \left(\prod_{a \neq b} \dd \mb_{ab}\right)  \prod_{a<b} \braket{\mb_{ab}}{\nb_{ab}}_{(1-c_{\gamma}) j_{ab}}\\
\hphantom{A_v =}{}
\times
  \braket{\mb_{ba}}{\nb_{ba}}_{(1-c_{\gamma}) j_{ab}} \bra{- \mb_{ab}} g_a^{-1} g_b \ket{\mb_{ba}}_{j_{ab}} \overline{\bra{-\mb_{ab}} {g'}_a^{-1} g'_b \ket{\mb_{ba}}}_{c_\gamma j_{ab}},
\end{gather*}
which is related to the form most often found in the literature by parametrising with $k = (1-c_\gamma) j$ instead.

\subsubsection{Asymptotics}

The geometricity of the amplitudes in the asymptotic regime that was the starting point of our construction, extends to the evaluation of the action on the critical points. If our boundary data  is the boundary data of a geometric 4-simplex $\sigma^4$ the phase is explicitly given by the Regge action of gravity
\[
S_{\rm R}
\big(\sigma^4\big) = \sum_{\Delta_{ab}} \Theta_{ab} |\Delta_{ab}|.
\]
Here $\Delta_{ab}$ are the triangles of $\sigma^4$, $|\Delta_{ab}|$ their areas, and $\Theta_{ab}$ the internal dihedral angle at $\Delta_{ab}$. On geometric boundary data the asymptotics of all models are of the form:
\begin{gather*}
A_v (\nb, \lambda j) \sim \left( \frac{2 \pi}{ \lambda} \right)^{12}\left( \sum_{\epsilon, \epsilon' = \pm 1} \frac{e^{i \lambda (\epsilon \pm \epsilon' c_\gamma) S_{\rm R}(\sigma^4)}}{N_{\epsilon \epsilon'}} + O\left(\frac1\lambda\right)\right),
\end{gather*}
where the $\pm$ in the exponent, and the $N_{\epsilon \epsilon'}$ depend on the precise model chosen, and $S_{\rm R}(\sigma^4)$ is the Regge action of the geometric 4-simplex with boundary geometry given by the tetrahedral outward normals~$j \nb$.

\subsection{Lorentzian}\label{sec-Lor}

For the Lorentzian theory the geometric constructions above are not immediately available. The decomposition into left and right sector is complicated by the fact that the Hodge operator in Lorentzian signature has eigenvalues~$\pm i$, thus it has no eigenstates in the real Lie algebra. Furthermore coherent bivectors are not readily available. However, there still is the equivalent of the diagonal subgroup. In fact the Lie algebra can be written as
\begin{gather*}
  [K_i,K_j]  =  \epsilon^{ijk} L_k,\qquad
[K_i, L_j]  =  - \epsilon^{ijk} K_k,\qquad
[L_i, L_j]  =  - \epsilon^{ijk} L_k.
\end{gather*}

Now the $K$, the generators of the boosts, still parametrize the coset space of the $\SU(2)$ subgroup generated by~$L$, that is, 3-dimensional hyperbolic space. The condition that the highest weight eigenvalue of a boost in a particular direction should be proportional to a rotation in the subgroup can be translated back into a representation theoretic criterion. To see how remember f\/irst that an irreducible representation of $\SL(2,\C)$ is labeled by two numbers, $k$ and $p$ where~$k$ is half integer and~$p$ is real positive. The representation $(p,k)$ decomposes into inf\/initely many representations of the $\SU(2)$ subgroup as such
\begin{gather*}
  V(p,k) = \bigoplus_{k' > k}V(k').
\end{gather*}

Therefore we can label coherent $\SU(2)$ states in $(p,k)$ through $\ket{j,\nb}$. Acting on these states we can require that $\nb \cdot K$ should be proportional to $\nb \cdot L$. This has a simple solution if we chose $k = j$. Then the eigenvalues of $K$ and $L$ in are simply $pkj/(j(j+1))$ and $j$. Thus we can achieve proportionality by choosing $p= \gamma (j+1)$.

This def\/ines an injection map along similar lines as $I_{\rm EPRL}$ in the Euclidean case
\[
I^{\rm Lor}_{\rm EPRL}(\ket{\nb}_j) = \ket{j,\nb}_{(\gamma (j+1), j)}.
\]

This way of deriving the injection map was f\/irst given in \cite{Ding:2010ye}. We again have the rule that the subgroup representation is inserted into the lowest weight representation of the group, the amplitude is simply given by
\begin{gather*}
A_v = \int_{\SL(2,\C)^{5}} \left(\prod_{c=1,\dots,5} \dd g_c \right) \prod_{a<b} \bra{-\nb_{ab},j_{ab}} g_a^{-1} g_b \ket{\nb_{ba}, j_{ab}}_{(\gamma j_{ab} + 1, j_{ab})} \delta(g_5).
\end{gather*}

This form of the amplitude is very abstract and dif\/f\/icult to calculate with, it can however be written very explicitly by using the formulation of $\SL(2,\C)$ representations as functions $f(z)$  with specif\/ic scaling behaviour on $z \in \C^2$. A group element $g \in \SL(2,\C)$ acts by $g f(z) = f(g^\dagger z)$. The formulation was given in~\cite{Barrett:2009mw}. In this language the map $I_{\rm EPRL}^{\rm Lor}$ takes the simple form
\begin{gather*}
I_{\rm EPRL}^{\rm Lor} (\ket{\nb}_j) (z) = \la z|z\ra^{-1-i\gamma(j+1)-j}_\frac12 \la z|\nb_{ab}\ra_{j}.
\end{gather*}

\looseness=-1
The inner products are in the $\SU(2)$ representations. We see immediately that the transformation behavior under the $\SU(2)$ subgroup, which acts unitarily and for which the f\/irst term is constant, is indeed as expected. Remarkably the f\/irst factor is simply the integration kernel for the Lorentzian amplitude of Barrett and Crane~\cite{Barrett2000}. Thus we can interpret the Lorentzian amplitude as combining the geometry of~$\SU(2)$ coherent states with that of the Barrett--Crane amplitude.

\subsection{Holomorphic intertwiners} \label{sec:Furthers}

An important signif\/icant extension and ref\/inement of the above construction was developed by Dupuis and Livine in the series of papers \cite{Dupuis:2010iq,Dupuis:2011dh,Dupuis:2011fz,Dupuis:2011wy, Dupuis:2012vp}.

Instead of using the coherent intertwiners of \cite{Livine:2007vk} they use a holomorphic version. That is, instead of our intertwiners $\iota(\nb)$, \eqref{eq-coherentInts}, they construct coherent intertwiners directly from spinors~$|z\rangle$ in~$\C^2$. This allows them to also bring in a vast set of tools from the study of the $\U(N)$ action on $\SU(2)$ intertwiners \cite{Borja:2011pd,Borja:2010rc,Freidel:2009ck,Freidel:2010tt}.

One particular choice of coherent intertwiner they discus, which they denote $||\{z_i\}\ra$, as it depends on a set of spinors $\{z_i\}$, is given by
\begin{gather*}
\iota(z_i) = ||\{z_i\}\ra = \int_{\SU(2)} \dd g \sum_{j_i} \bigotimes_i \frac1{\sqrt{2j_i !}} g |z_i\rangle^{2j_i}.
\end{gather*}

In the papers many other equivalent forms of this intertwiner are given. Note that every element $|z\rangle$ is an unnormalized coherent state in $j=\frac12$ for some $\nb$, and that $|\nb\rangle_{j} = |\nb\rangle_{\frac12}^{2j}$. This state has support in all representations $j_i$. The asymptotic behaviour is now given in terms of a joint scaling of the $|z_i\rangle$. These intertwiners still retain the geometric interpretation of the $\nb$, with the length square of the spinors $|z_i\rangle$ now playing the role of the spins~$j_i$.

The construction of a vertex amplitude by contraction in the sense of~\eqref{eq-ContractionPattern} from this gives us an alternative formulation of BF theory, and we can again immediately construct a theory with a~geometric sector by doubling and identifying $z^+_i = \sqrt{c_\gamma} z^-_i$, which ensures that the length squares have the same relation as we expect from the $j^\pm_i$. From the asymptotic geometric interpretation of the spinors we expect that this corresponds to the construction $\nb^+ = \nb^-$, and that $\nb^+ = - \nb^-$ can be achieved by using the antilinear, $\SU(2)$ covariant map $J$ that which has the property that $J | \nb \rangle = | - \nb \rangle$. Thus for that case we would have $z^+_i = J \sqrt{c_\gamma} z^-_i$.

\looseness=-1
While they encode the same asymptotic geometry, these intertwiners arise by implementing this geometry conditions using constraints that commute, avoiding the construction in terms of weakly implementing noncommuting constraints we will encounter in Section~\ref{sec-Principles}. Note that this does not neccesarily mean that the resulting quantum theory is dif\/ferent. A~further advantage of these intertwiners is that they are holomorphic in the spinors~$|z_i\rangle$, and their functional form is simpler. This allows us to carry computations much further by using spinor techniques, e.g.~\cite{Borja:2011pd,Livine:2011gp,Livine:2011vk,Livine:2011zz} and references therein. They are in fact closely related to harmonic oscillator coherent states.

\subsection[$B$-field quantization]{$\boldsymbol{B}$-f\/ield quantization}

Another important recent development advocated in \cite{Baratin:2010wi,Baratin:2011tx,Baratin:2011hp}, is to change the perspective on the quantization of BF theory that we use to one which makes the $B$-f\/ields explicit.

The way we have been writing down the BF partition function above has been in terms of states on the group. An alternative is to write it in terms of the quantization of the Lie algebra directly. This leads to the notion of non-commutative plane waves. The relationship to the formalism above is established by use of a non-commutative Fourier transform that respects the group structure \cite{Freidel:2005bb,Freidel:2005ec,Joung:2008mr,Livine:2008hz}. Following \cite{Baratin:2010wi,Baratin:2011tx,Baratin:2011hp}, we can then decompose~$\delta(g)$ into non commutative plane waves.

\looseness=-1
Thus the simplicity constraints can then be implemented on the Lie algebra elements directly. The simplest way of doing so reproduces the Barrett--Crane model, however a new class of amplitudes that incorporate $\gamma$ can also be def\/ined. For details we refer the reader to~\cite{Baratin:2011hp}. This has the advantage that, while in the above the geometric construction only f\/ixes the asymptotics of the amplitude, here the geometry is manifest, albeit in a non-commutative sense, in the quantum regime as well. The ambiguities in extending to the non-asymptotic regime here are given explicitly in terms of the quantization map, or choice of non-commutative fourier transform, and thus appear as genuine quantization ambiguities.

\section{Spin foam vertex as a Feynman interaction vertex for GFT}\label{sec-Principles}

Originally, Reisenberger and Rovelli proposed the spinfoam framework as a technique to def\/ine the covariant dynamics of Loop Quantum Gravity (LQG) \cite{Reisenberger1997a}. Soon after, Markopoulou and Smolin presented a~list of requirements the spinfoam vertex amplitude has to satisfy in order to provide the causal evolution of spin-network states~\cite{Markopoulou:1997wi}. In this section we develop this line of reasoning and present a derivation of the spinfoam vertex amplitude starting from few general principles, taking the form of the geometricity constraints~$C(B)$ of~\cite{Ding:2010ye} as a key input. The construction focuses on the interaction of quantum grains of space, and is complementary to the one described in the previous section. This is most naturally viewed in the context of a f\/ield theory that generates these vertex amplitudes and their gluing into Feynman diagrams. For BF theory such f\/ield theories were def\/ined by Boulatov~\cite{Boulatov:1992vp} and Ooguri~\cite{Ooguri1992c}. The corresponding f\/ield theory for Barrett--Crane was f\/irst given in~\cite{DePietri:1999bx}, the amplitude for the EPRL-FK type models we are considering here was f\/irst given in~\cite{Geloun:2010vj}, for the Baratin--Oriti model it is in~\cite{Baratin:2010wi,Baratin:2011tx,Baratin:2011hp}.

In quantum electrodynamics, the Feynman interaction vertex can be determined by identi\-fying the states of the theory (photons and electrons) and requiring that their interaction is local in spacetime and invariant under Lorentz transformations. The spinfoam vertex amplitudes can be characterized by the following requirements:
\begin{itemize}\itemsep=0pt
\item[i.] {\bf Degrees of freedom}. The degrees of freedom are those of $\SU(2)$ BF. They are interpreted as quanta of the 3-geometry, as in Loop Quantum Gravity.
\item[ii.] {\bf Lorentz covariance}. These quanta transform under Lorentz transformations covariantly. They maintain their interpretation as quanta of 3-geometries, but now in an ambient 4-dimensional space. It is here that the geometricity constraints enter, by connecting the 3- and 4-geometries.
\item[iii.] {\bf Interaction}. These quanta can interact ``locally'' in the sense of \cite{Krajewski2010a}\footnote{As there is no space-time here, the notion of locality is fundamentally dif\/ferent from that of ordinary space time physics. The meaning of locality in~\cite{Krajewski2010a} is deeply related to the fact that the theory is formulated on a~2-complex and not on a~higher-dimensional combinatorial object.} changing in number. Their interaction is invariant under Lorentz transformations.
\end{itemize}
Let us elaborate on these three points.

\subsection{Degrees of freedom: quantum polyhedra}
In Loop Quantum Gravity, the state of a quantum grain of space is described by a $\SU(2)$ intertwiner $|i\rangle$,
\begin{gather*}
|i\rangle=\sum_{m_f}i^{m_1\cdots m_F}\,|j_1,m_1\rangle\cdots |j_F,m_F\rangle.
\end{gather*}
It can be understood as a quantum polyhedron in 3d Euclidean space $\R^3$,~\cite{Bianchi:2010gc}. The vector normal to a face of the polyhedron corresponds to the $\SU(2)$ generators $\vec{L}$. Its state is $|j,m\rangle$. At the classical level, the normals to the faces of the polyhedron sum to zero. This corresponds to the $\SU(2)$-invariance condition $(\vec{L}_1+\cdots+\vec{L}_F)|i\rangle=0$ at the quantum level. These are the degrees of freedom of the theory.

Note that these quanta do not form a continuous 3-geometry, nor even a cellular decomposition of 3-space. The discrepancy between the space of geometries and the space of polyhedra, along with constraints to reduce the latter to the former was discussed in~\cite{Dittrich2008,Dittrich2010a,Dittrich2011}. For more on this in this special issue see also~\cite{Dupuis:2012yw}.

\subsection{Lorentz covariance: quantum polyhedra in 4d Minkowski space}

As the LQG degrees of freedom don't naturally have an action of the Lorentz group on them we need to inject them into representations of the Lorentz group. A~quantum polyhedron transforming covariantly in 4d Minkowski space is obtained by identifying the rotation group $\SU(2)$ with the little group of the the Lorentz group $\SL(2,{\mathbb C})$ that leaves invariant a time-like vector $t^I$. In terms of unitary representations, we have the map (see Appendix~\ref{app:reps})
\begin{gather*}
Y:\ |j_f,m_f\rangle\mapsto |(p_f,k_f);j_f,m_f\rangle .
\end{gather*}
Similarly, for the interwiner we have{\samepage
\begin{gather*}
|Yi\,\rangle=\sum_{m_f}i^{m_1\cdots m_F}\,|(p_1,k_1); j_1,m_1\rangle\cdots |(p_F,k_F); j_F,m_F\rangle.
\end{gather*}
Our task now is to identify the relevant $\SL(2,{\mathbb C})$ representation $(p_f,k_f)$.}

The shape of a 3d Euclidean polyhedron embedded in 4d Minkowski space can be described using as variables a timelike vector $t^I$ and $F$ spacelike vectors $A_f^I$ satisfying the conditions
\begin{gather*}
t_I A_f^I=0  \qquad \text{(rest frame)} ,
\end{gather*}
and
\begin{gather*}
\sum_f A_f^I=0  \qquad \text{(closure)} .
\end{gather*}
The time-like vector $t^I$ identif\/ies the rest frame of the polyhedron. The space-like vectors $A_f^I$ describe the 3-normals to the N faces of the polyhedron, their norm being the area of the face. The fact that, up to rotations and translations, there exists a unique polyhedron with this data is a consequence of Minkowski theorem \cite{Bianchi:2010gc}. The orientation of a face of the polyhedron in 4d is described by the bivector  $\Sigma_f^{IJ}$,
\begin{gather}
\Sigma_f^{IJ}={\epsilon^{IJ}}_{KL}A_f^K t^L .
\label{eq:sigmaAt}
\end{gather}
This bivector is \emph{simple}, meaning that it is of the form $\Sigma^{IJ}_f=e_1^I e_2^J-e_1^J e_2^I$, where $e_1^I$ and $e_2^I$ are two vectors that span the face $f$. In the following we describe the shape of a polyhedron in 4d using an equivalent set of variables: we use as data the timelike vector $t^I$ and a set of $F$ bivectors $\Sigma^{IJ}_f$ satisfying the following constraints
\begin{gather}
R_f^I\equiv t_N \Sigma^{NI}_f = 0\qquad  \text{(rest frame')} ,\label{eq:restframe}
\end{gather}
and
\begin{gather}
\sum_f t_N \frac{1}{2}{\epsilon^{NI}}_{JK}  \Sigma_f^{JK} = 0   \qquad \text{(closure')} .\label{eq:4-closure}
\end{gather}
The condition (\ref{eq:restframe}) codes the rest frame of the polyhedron, (\ref{eq:4-closure}) the closure condition. The area-vector is now a derived quantity and given by $A_f^I=t_N \frac{1}{2}{\epsilon^{NI}}_{JK}  \Sigma_f^{JK}$.

  The rest-frame condition (\ref{eq:restframe}) can be imposed at the quantum level requiring that the following conditions hold
\begin{gather*}
 \langle Y_\gamma\, i'|\,\hat{R}_f^{I} |Y_\gamma  i\rangle=0\quad \forall\, f\qquad   \text{(quantum rest frame)}\qquad
 \text{and smallest dispersion} \ \Delta R_f^I .
\end{gather*}
These two conditions identify the Hilbert space of quantum polyhedra in 4d Minkowski space, $\mathcal{H}=\operatorname{Im} Y_\gamma$. Notice that the rest frame condition is imposed weakly on this Hilbert space\footnote{The commutator of rest-frame operators is $[\hat{R}^I,\hat{R}^J]=-i\frac{\gamma}{1+\gamma^2} t_N {\epsilon^{NIJ}}_K (\gamma \hat{R}^K+\hat{A}^K)$, where $\hat{A}^I$ is the area operator. As a result, imposing the rest-frame condition strongly, $\hat{R}^I_f=0$, would require vanishing areas.}. To determine $Y_\gamma$, we need to express the operator $\hat{\Sigma}_f^{IJ}$ in terms of generators $J_f^{IJ}$ of the Lorentz group. Here enters the key relation between LQG, BF theory, and the action (\ref{eq-HolAction}) of general relativity in terms of constrained $B$-f\/ields. In LQG, the momentum conjugated to the Ashtekar connection acts on states as a generator $\vec{L}$ of $\SU(2)$ transformations. Similarly, in quantum BF theory, the momentum conjugated to the Lorentz connection is the $B$-f\/ield (as can be read from the action~(\ref{eq-BFAction})). It acts on states as a generator~$J^{IJ}$ of the Lorentz group. Given these preliminaries, we can now look for a quantum version of the classical relation~(\ref{eq:simpleB}) between the~$B^{IJ}$ and the metric two-form~$e^I\wedge e^J$. The proposed quantum version of~(\ref{eq:simpleB}) in terms of the operators~$\hat{\Sigma}_f^{IJ}$ and~$J_f^{IJ}$ is
\begin{gather*}
J_f^{IJ}=\frac{1}{2}{\epsilon^{IJ}}_{KL} \hat{\Sigma}_f^{KL} + \frac{1}{\gamma} \hat{\Sigma}_f^{IJ} .
\end{gather*}
This expression can be inverted to f\/ind $\hat{\Sigma}_f^{IJ}$,
\begin{gather*}
\hat{\Sigma}_f^{IJ}=\frac{\gamma}{1+\gamma^2}\left( J_f^{IJ} - \gamma \frac{1}{2}{\epsilon^{IJ}}_{KL} J_f^{KL}\right).
\end{gather*}
We immediately see that the quantum rest-frame operator is given by
\begin{gather*}
\hat{R}_f^{I}=t_N \hat{\Sigma}_f^{NI}=\frac{\gamma}{1+\gamma^2}\big( K_f^I-\gamma L_f^I\big)
\end{gather*}
where $L^I=t_N \frac{1}{2}{\epsilon^{NI}}_{JK}  J^{JK}$ and $K^I=t_N J^{NI}$ are the generators of rotations and boosts discussed in the Appendix~\ref{app:reps}.
The requirement that on the Hilbert space $\mathcal{H}=\operatorname{Im} Y_\gamma$ the matrix elements of $\hat{R}_f^{I}$ vanish coincides with the weak imposition of the linear simplicity constraint
\begin{gather*}
\langle Y_\gamma i'|\,  K_f^I-\gamma L_f^I \, |  Y_\gamma  i\rangle=0\qquad \forall\, f
\end{gather*}
discussed in  \cite{Ding:2010ye}. It select representations $p_f$ and $k_f$ of the form \cite{Ding:2010ye,Ding:2010fw}
\begin{gather*}
p=\gamma j \frac{j+1}{j-r} ,\qquad  k=j-r ,
\end{gather*}
where\looseness=-1  $r$ is an integer in $[0,j]$. The dispersion $\Delta R_f^I$ can be computed, and it attains its smallest value at $r=0$. Therefore, the relevant representations are $p_f=\gamma(j_f+1)$, $k_f=j_f$, and the map~$Y_\gamma$~is
\begin{gather*}
Y_\gamma |j,m\rangle = |(\gamma(j+1),j);j,m\rangle .
\end{gather*}
Now we look at the geometry described by the area 4-vectors $A_f^I$. The area operator is here a~derived quantity
\begin{gather*}
\hat{A}_f^I=t_N \frac{1}{2}{\epsilon^{NI}}_{JK}  \hat{\Sigma}_f^{JK}=\frac{\gamma}{1+\gamma^2}\big(L_f^I+\gamma K_f^I\big)
\end{gather*}
and on the rest-frame Hilbert space $\mathcal{H}=\operatorname{Im}Y_\gamma$, it reduces to
\begin{gather*}
\hat{A}_f^I\approx \gamma L_f^I ,
\end{gather*}
and matches with the area operator of LQG.

This completes the description of a quantum polyhedron at rest, $|Y_\gamma\, i\rangle$. The state now transforms covariantly under the appropriate symmetry group,
\begin{gather*}
\U(G)\,|Y_\gamma  i\rangle ,
\end{gather*}
where $\U(G)$ is the unitary representation of a Lorentz transformation~$G$.

\subsection{Interaction}
We assume that $M$ quantum polyhedra can decay in $N$ quantum polyhedra via a ``local'' interaction. The transition amplitude depends linearly on the state of each of the quantum polyhedra,
\begin{gather*}
A: \ \bigotimes_{e=1}^M U(G_e)\, |Y_\gamma  i_e\rangle\to\bigotimes_{e=M+1}^{M+N} U(G_e)\, |Y_\gamma  i_e\rangle .
\end{gather*}
We assume that the transition amplitude is invariant under Lorentz transformations applied to each polyhedron, i.e.\ it does not depend on~$G_e$. Thus we need to contract it with an invariant. Locality in the sense of~\cite{Krajewski2010a} can be expressed as the condition that this invariant is the direct product of copies of the invariant bilinear form. In the language of~\cite{Krajewski2010a} this is called simpliciality.

We then obtain interactions of the form
\begin{gather*}
A_0(i_1,\dots ,i_N)=\int_{\SL(2,\C)}\prod_{e=2}^N dG_e  \Bigg\{\bigotimes_{e=1}^N U(G_e) |Y_\gamma  i_e\rangle\Bigg\}
\end{gather*}
where again $\{~\}$ is the contraction with the bilinear invariant form. This def\/ines the transition amplitude from $0$ to $N$ quanta\footnote{Notice that the integration is over $N-1$ copies of $\SL(2,\C)$. This is enough to make the amplitude invariant under Lorentz transformations, and an extra integration would simply give the inf\/inite volume of the Lorentz group~\cite{Crane:2001qk,Engle:2008ev}. In the Euclidean case, the volume of $\SO(4)$ is f\/inite and this point can be disregarded. In that case the amplitude can be simply written as a contraction of $\SO(4)$ intertwiners as $A_0(i_1,\dots ,i_N)=\big\{\otimes_{e\in v}\mathcal{I}_\gamma(i_e)\big\}$,  where $\mathcal{I}_\gamma(i_e)=\int dG_e\, U(G_e)|Y_{\gamma} i_e\rangle$.}. The transition amplitude from $M$ \emph{ingoing}  to $N$ \emph{outgoing} quanta can be def\/ined introducing the time-reversal operator as in~\cite{Bianchi:2011hp}. It is given by
\begin{gather*}
A(i_1,\dots ,i_M\to i_{M+1},\dots ,i_{M+N})=A_0(\mathcal{T} i_1,\dots ,\mathcal{T} i_M,  i_{M+1},\dots , i_{M+N}) .
\end{gather*}
The ef\/fect of the time-reversal operators $\mathcal{T}$ is to multiply the amplitude $A_0(i_1,\dots ,i_{M+N})$ by a~spin-dependent sign $(-1)^{\sum'_f j_f}$, with the sum $\sum'_f$ restricted to faces $f$ shared by an ingoing and an outgoing quantum polyhedron.

\looseness=-1
The construction of the spinfoam amplitude based on the dynamics of LQG naturally respects its structure. In particular there is no restriction to $4$-valent spin-network graph and simplicial decompositions of the manifold. As  a result, it generalizes simplicial spin-foams along the lines discussed in \cite{Kaminski2010,Kaminski2010a,Kaminski2011} and in~\cite{Ding:2010fw}. Remarkably, when restricting attention to transition amplitudes for $5$ quantum tetrahedra, the result of this construction coincides with the vertex amplitude of section \ref{sec-GeometricConstruction}, the one obtained imposing geometricity on BF theory. In particular, all the results about the geometricity of the $4$-dimensional conf\/iguration apply, and the asymptotics of the amplitude for semiclassical tetrahedra satisfying the Regge geometricity conditions reproduces the cosine of the Regge action.

\section{Discussion and open issues}\label{sec-OpenIssues}
Spin Foam vertex amplitudes constitute an attempt to code the quantum gravitational dynamics in an elementary group theoretical object. They provide a 4-dimensional generalization of the Ponzano--Regge model for 3d quantum gravity \cite{ponzanoregge}. In this paper we have presented two distinct derivations of the spin foam vertex amplitude.

\looseness=-1
The f\/irst is based on geometricity imposed on quantum BF theory. The advantage of this formulation is that its geometric content is available from the beginning. It identif\/ies the properties necessary to obtain the Regge action as a phase in the semiclassical limit. It is possible to do so in such a way to obtain the boundary Hilbert space of LQG restricted to a simplicial graph.

The second construction is an attempt to extract from the construction through constraints a set of construction principles to determine the dynamics of the LQG degrees of freedom. As a~result, the LQG space of states is a starting point. Lorentz invariance of the interaction and an analogue of the constraints of the classical theory largely determine the structure of the vertex amplitude. In this approach, the asymptotic geometry results can be seen as a test for the analogous constraints chosen.

\looseness=-1
The geometricity results extend to the actual evaluation of these amplitudes in the asymptotic regime. There the phase of the various amplitudes described here reduces to the Regge action. Thus it is reasonable to take these amplitudes as starting points for lattice quantisation of gravi\-ta\-tional theories. However, we do not have any geometricity results for arbitrary 2-complexes and boundary spin networks that are not simplicial. We see that in order for these lattice models to be compatible with the LQG boundary state space, a number of trade of\/fs have to be made, the loss of geometricity when going to generic graphs being maybe the most severe one.

Note further that, despite the appearance of the Regge action, the calculations based on these amplitudes so far do not test the dynamics of the theory. To do so it would be necessary to study the sum over spins directly. Indeed, initial results in this direction did indicate that the dynamics might suf\/fer from a f\/latness problem \cite{Bonzom:2009hw, Hellmann:2012kz}, in the sense that only a discrete set of curvatures are allowed.

Of particular worry is that next to the geometric sector, the topological $\SU(2)$ BF sector remains in the theory. In the Barrett--Crane model it were these topological solutions that led to the dominant behaviour of the vertex amplitude \cite{Baez2002e}\footnote{Note though that at least in the Lorentzian amplitudes discussed above this sector appears cleanly separated from the geometric one.}. Without studying the dynamics of the theory it remains unclear whether their presence continues to dominate the geometric sector.

An outstanding challenge for the construction of interesting amplitudes is to understand how to eliminate this sector. This sector does appear already in the classical implementations based on simplicity constraints \cite{Capovilla:1989ac,Capovilla:1991qb,Reisenberger1998}. Dittrich and collaborators have given a formulation of constraints that eliminate the non-geometric sector entirely for discrete classical data in~\cite{Dittrich2011,Dittrich2010a,Dittrich2008}, but no good implementation of these constraints in the spin foam formalism is known to date.

A related issue that is shared by the Ponzano--Regge model, too, is that we obtain a sum over orientations in the asymptotic regime. There have been several attempts to modify the amplitudes in order to select only one of these sectors, see~\cite{Engle:2012yg,Rovelli:2012yy,Engle2011e,Livine2003,Livine2003a}.

Finally, it is necessary to understand what, if any, continuum behaviour arises from the discretized theory. This remains by far the most ununderstood and important challenge facing the spin foam approach to quantum gravity. In order to study the large scale, or emergent behaviour of the theory we need to understand how the vertex amplitude changes under coarse graining, and be able to def\/ine, if possible, ef\/fective vertex amplitudes. Initial steps towards studying what aspects of the vertex amplitude will turn out to be relevant at larger scales have been undertaken \cite{Bahr2009,Bahr2009a,Bahr2009b,Bahr2010,Bahr2011b}, but the situation remains highly unclear and an open challenge.

In the context of the group f\/ield theory perspective the situation is somewhat better. A~class of f\/ield theories, which are however only very loosely related to the spin foam amplitudes consi\-de\-red here, have been shown very recently to be indeed renormalizable \cite{BenGeloun:2011xu,Geloun:2011cy,Geloun:2010nw,Geloun:2009pe,Carrozza:2012uv,Rivasseau:2011xg,Tanasa:2011ur}. Whether this will extend to models incorporating the geometric f\/lavours we considered here remains an open question under active consideration.
Here it is clear that in both cases the gluing of the amplitudes, either in a lattice using face amplitudes, or into a Feynman diagram using propagators, will be equally important to the def\/inition of the vertex itself.

As an attempt to address the problem of quantum gravity, spin foams vertex amplitudes should thus be understood as elementary building blocks providing a starting point for def\/ining a theory. From this perspective, they should be evaluated only on the basis of their properties and on the consequences that follow from them, and the way they address the open issues discussed above. The value of the heuristic derivations present in the literature and described in this paper is that they lead to a specif\/ic proposal for such building blocks, and can provide a~valuable guide in the process of building a theory of spin foam quantum gravity.

\appendix

\section{Unitary representations of the rotation and the Lorentz group}\label{app:reps}

Unitary representations of the group $\SU(2)$ on a Hilbert space $V$ are generated by three hermitian operators $L^i$, $i=1,2,3$ obeying the commutation relations\footnote{In the following we use also the vector notation $\vec{x}=x^i$ and $\vec{x}\cdot \vec{y}=\delta_{ij}x^i y^j$.}
\begin{gather*}
[L^i,L^j]=i{\epsilon^{ij}}_k L^k .
\end{gather*}
Irreducible representations $V^{(j)}$ are labeled by an half-integer $j=0,\frac{1}{2},1,\ldots$, the \emph{spin}, and are f\/inite-dimensional $\dim V^{(j)}=2j+1$. We follow the standard notation and call
\begin{gather*}
|j,m\rangle\in V^{(j)}
\end{gather*}
an orthonormal basis of simultaneous eigenstates of the Casimir operator $\vec{L}^2$  and of a component $L_z=\vec{L}\cdot\vec{z}$. The eigenvalues are $j(j+1)$ and $m=-j,\ldots,+j$ respectively.

Unitary representations of the group $\SL(2,\C)$ on a Hilbert space $\mathcal{V}$ are inf\/inite-dimensional and are generated by six hermitian operators\footnote{In the following, $\eta_{IJ}$ is the Minkowski metric with signature $-+++$, and $\epsilon^{IJKL}$ the Levi-Civita tensor ($\epsilon^{0123}=+1$).} $J^{IJ}=-J^{JI}$, $I,J=0,1,2,3$.  Let us introduce a unit time-like vector $t^I$, and def\/ine the generator of Lorentz transformations that leave $t^I$ invariant (rotations) as
\begin{gather*}
L^I=\frac{1}{2}{\epsilon^I}_{JKL}J^{JK}t^L ,
\end{gather*}
and the generator of boosts of $t^I$ as
\begin{gather*}
K^I=J^{IJ}t_J .
\end{gather*}
Notice that $t_I L^I=t_I K^I=0$ so that, in coordinates such that $t^I=(1,0,0,0)$, we have $L^I=(0,L^i)$ and $K^I=(0,K^i)$. These generators of $\SL(2,\C)$ obey the following commutation relations
\begin{gather*}
[L^i,L^j] =+i {\epsilon^{ij}}_{k}\,L^k ,\qquad
[L^i,K^j] =+i {\epsilon^{ij}}_{k}\,K^k ,\qquad
[K^i,K^j] =-i {\epsilon^{ij}}_{k}\,L^k .
\end{gather*}
Unitary irreducible representations $\mathcal{V}^{(p,j)}$ of $\SL(2,\C)$  (the principal series,~\cite{Ruhl}) are labeled by a real number $p$ and a half-integer $j$. As $\SU(2)$ is a subgroup of $\SL(2,\C)$, they are also unitary representations of $\SU(2)$, but reducible. In particular they decompose into irreducible representations as follows
\begin{gather}
\mathcal{V}^{(p, j)}= V^{(j)}\oplus V^{(j+1)}\oplus V^{(j+2)}\oplus \cdots .
\label{eq:Vjj}
\end{gather}
In $\mathcal{V}^{(p,j)}$, the two invariant Casimir operators $C_1$ and $C_2$ have eigenvalue
\begin{gather*}
C_1 =\frac{1}{2}J_{IJ}J^{IJ} = \vec{K}^2-\vec{L}^2 = p^2-j^2+1 ,\qquad
C_2 =\frac{1}{8}\epsilon_{IJKL}J^{IJ}J^{KL} = \vec{K}\cdot\vec{L} = p j .
\end{gather*}
We denote by{\samepage
\begin{gather*}
|(p,j); j',m\rangle\in\mathcal{V}^{(p,j)}
\end{gather*}
with $j'\geq j$ an orthonormal basis of simultaneous eigenstates of $\vec{L}^2$ and $L_z=L^I z_I$, with $t^Iz_I=0$. }

The groups $\SU(2)$ and $\SL(2,C)$ are respectively the double cover of the rotation group $\SO(3)$ and of the part of the Lorentz group connected to the identity, $\SO^{\uparrow}(3,1)$. In the text, we refer loosely to them as the rotation and the Lorentz group.

The decomposition (\ref{eq:Vjj}) allows to identify the vector $|j'm\rangle$ that transforms under the representation~$j'$ of the rotation group with the vector $|(p,j); j',m\rangle$ that transforms in the representation~$j'$ of the little group of the Lorentz group that leaves the time-like vector $t^I$ invariant. In spinfoams, the map $Y_{\gamma}$ that identif\/ies the representation $V^{(j)}$ of $\SU(2)$ with the lowest-spin block in the decomposition (\ref{eq:Vjj}) plays a special role. More explicitly, calling~$\gamma$ the ratio between~$p$ and~$j+1$, the map is def\/ined by
\begin{gather*}
Y_{\gamma}:  \ V^{(j)}\to  \mathcal{V}^{({ \gamma (j+1)}, j)}, \qquad
 |j,m\rangle\mapsto |({ \gamma (j+1)},j); j,m\rangle .
\end{gather*}
This map has the following notable property: it identif\/ies a Hilbert space $\mathcal{V}_\gamma=\operatorname{Im}Y_\gamma$ where the matrix elements of the operator $K^i-\gamma L^i$ vanish,
\begin{gather*}
\langle({ \gamma (j+1)},j); j,m'|\,K^i-\gamma L^i\,|({ \gamma (j+1)},j); j,m''\rangle = 0 .
\end{gather*}
This is the linear simplicity constraint discussed in Section~\ref{sec-Principles}.

\pdfbookmark[1]{References}{ref}
\LastPageEnding

\end{document}